# COUPLING-DEPENDENT SWITCHING BETWEEN SYNCHRONIZATION REGIMES IN FEEDBACK TIME DELAYS LASER SYSTEMS

E.M.Shahverdiev[1], P.A.Bayramov[1] and K.A.Shore[2]

[1]Institute of Physics, H. Javid Avenue,33, Baku, AZ1143, Azerbaijan

[2]School of Electronic Engineering, Bangor University, Dean St.,Bangor, LL57 1UT, Wales, UK

## ABSTRACT

We study a model for two lasers that are mutually coupled opto-electronically by modulating the pump of one laser with the intensity deviations from the steady-state output of the other. Such a model is analogous to the equations describing the spread of certain diseases such as dengue epidemics in human populations coupled by migration, transportation, etc. In this report we consider the possibility of both in-phase (complete) and anti-phase (inverse) synchronization between the above-mentioned laser models. Depending on the coupling rates between the systems a transition from the in-phase (complete) synchronization to the anti-phase (inverse) synchronization might occur. The results are important for the disruption of the spread of the certain infectious diseases in human populations.

## INTRODUCTION

Synchronization of chaos refers to a process wherein two (or many) chaotic systems (either equivalent or nonequivalent) adjust a given property of their motion to a common behavior due to a coupling or to a forcing (periodical or noisy)[1]. Chaos synchronization is relevant especially in secure communication, physiology, nonlinear optics and fluid dynamics, etc.[1-2].

In this paper we study a model for two lasers that are mutually coupled opto-electronically by modulating the pump of one laser with the intensity deviations from the steady-state output of the other. As shown in [3], such a model is analogous to the equations describing the spread of certain diseases such as dengue epidemics in human populations coupled by migration, transportation, etc. In-phase oscillations (complete synchronization) between such epidemic models, which favors the spread of diseases are considered in [4]. In this report we consider the possibility of both in-phase (complete) and anti-phase (inverse) synchronization between the above-mentioned laser models. Changing the coupling rates between the systems we find that the possibility of the transition from the in-phase(complete) synchronization to the anti-phase (inverse) synchronization might occur. The results can be important for the disruption of the spread of the certain infectious diseases in human populations and chaos based secure communication.

## ORIGINAL LASER AND EPIDEMIC MODELS

Following [3] we consider a standard semiconductor laser model with the following intensity and carrier number dynamics

$$\frac{dI}{dt} = g\, I\, N - \tau_p^{-1} I \qquad (1)$$

$$\frac{dN}{dt} = P - g\, I\, N - \tau_c^{-1} N \qquad (2)$$

Here I is the photon number and describes fast dynamical processes; N-slow dynamical variable is the carrier number; Source term P is the pump rate; g is the gain coefficient responsible for the nonlinear coupling between I and P; $\tau_p$ is the photon lifetime; $\tau_c$ is the carrier lifetime.

The equivalence of the laser model (1) and (2) to the following epidemic model (3) and (4) describing the spread of certain diseases such as dengue epidemics in human populations coupled by migration, transportation, etc. is evident:

$$\frac{dI_e}{dt} = (\alpha(\mu+\gamma))\, k\, I_e S - (\mu+\alpha) I_e \qquad (3)$$

$$\frac{dS}{dt} = \nu - \mu S - k\, I_e S \qquad (4)$$

Here $I_e$ is the infective population number and equivalent to the laser photon number I in (1) and (2); S is the number of population susceptible to the infection and equivalent to the carrier number N in (1) and (2); $\nu$-susceptible input rate is equivalent to P - the pump rate in (1) and (2);k-contact rate is equivalent to g - the gain coefficient in (1) and (2); $\mu$-death rate; $\alpha^{-1}$-average latent period; $\gamma^{-1}$-average infectious period.

DIMENSIONLESS LASER MODEL

By introducing new variables J,D and $t_1$

$$I = I_0 J,\ N = N_0 D,\ t = t_0 t_1 \qquad (5)$$

with

$$t_0 = \tau_p,\ N_0 = \tau_p^{-1} g^{-1},\ I_0 = \tau_c^{-1} g^{-1} \qquad (6)$$

and

$$\varepsilon^2 = \tau_p \tau_c^{-1},\ A = P N_0^{-1} \tau_c \qquad (7)$$

The original laser model (1) and (2) can be rewritten as:

$$\frac{dJ}{dt_1} = (D-1)J \qquad (8)$$

$$\frac{dD}{dt_1} = \varepsilon^2 (A-(J+1)D) \qquad (9)$$

With a sequence of transformations (for details see, Ref. [5]), the system of Eqs. (8) and (9) can be rescaled to Eqs. (10) and (11)

$$\frac{dy}{dt} = x(1+y) \qquad (10)$$

$$\frac{dx}{dt} = -y - \varepsilon x(a + by) \qquad (11)$$

with

$y = (J-J_0)/J_0,\ x=(D-D_0)\ \varepsilon^{-1} J_0^{-0.5},\ a=(1+J_0)J_0^{-0.5};b=J_0^{0.5},\ t=\varepsilon J_0^{0.5} t_1$ (12)

$D_0=1$ and $J_0=A-1$ are the steady states for Eqs. (8) and (9).

Notice that y is the intensity fluctuation normalized about the steady state level.

TWO DELAY-COUPLED LASERS:

$$\frac{dy_1}{dt} = x_1(1+y_1) \tag{13}$$

$$\frac{dx_1}{dt} = - y_1 - \varepsilon\ x_1(a + by_1) - \varepsilon K_2\ y_2(t - \tau) \tag{14}$$

$$\frac{dy_2}{dt} = x_2(1+y_2) \tag{15}$$

$$\frac{dx_2}{dt} = - y_2 - \varepsilon\ x_2(a + by_2) - \varepsilon K_1\ y_1(t - \tau) \tag{16}$$

$K_1$ and $K_2$ are the coupling strengths.

NUMERICAL SIMULATIONS AND CONCLUSIONS

In the numerical simulations we choose the following values for the main parameters appropriate to semiconductor lasers [3]:

$\varepsilon=(0.001)^{0.5}$, a=2, b=2.33, $\tau=30$ and change the coupling constants $K_1$ and $K_2$.

Figure 1 shows the complete synchronization $y_1=y_2$ between mutually coupled laser systems described by Eqs.(13) and (14) and Eqs.(15) and (16) for the coupling constants $K_1=6$ and $K_2=6$.

In figure 2 we present the case of inverse synchronization $y_1=-y_2$ for the coupling constants $K_1=2.06$ and $K_2=2.09$ [3] .

Figure 3 depicts the dynamics of the correlation coefficient C between $y_1$ and $y_2$ whilst changing the coupling constant K2 and for fixed K1=2.06. With K1 fixed and K2 varying high values for the correlation coefficient C between y1 and y2 were not obtained (C being usually less than -0.60). More importantly for a wide range of the parameters for this case the dynamics was trivial, approaching the fixed state.

The numerical simulations clearly support the possibility of switching from complete synchronization to inverse complete synchronization. As mentioned above due to the multidisciplinary nature of the laser model considered in this report, the results might have important implications in preventing certain epidemic diseases from spreading.

In conclusion, we present the case of both complete and inverse chaos synchronization in one of opto-electronically coupled laser systems. As these laser systems are important for chaos based secure communication systems the results have certain implications for information processing in secure communication, e.g. message masking and decoding. Additionally such a laser model is analogous to the equations describing the spread of certain diseases such as dengue epidemics in human populations coupled by migration, transportation, etc. Switching between the complete and inverse synchronization regimes then could be helpful in the disruption of the spread of the certain infectious diseases.

ACKNOWLEDGEMENTS

This research was supported by a Marie Curie Action within the $6_{th}$ European Community Framework Programme Contract N: MIF2-CT-2007-039927-980065(Reintegration Phase).

FIGURE CAPTIONS

FIG.1. Numerical simulation of two delay coupled laser systems, Eqs.(9-12) for $\varepsilon$ =$(0.001)^{0.5}$, a=2, b=2.33, $K_1$=6, $K_2$=6, $\tau$ =30. Complete synchronization $y_1$=$y_2$: Time series of laser system $y_1$ intensity fluctuations is shown: C is the cross-correlation coefficient between the laser systems $y_1$ and $y_2$ intensity fluctuations. Dimensionless units.

FIG.2. Numerical simulation of two delay coupled laser systems, Eqs.(9-12) for $\varepsilon$ =$(0.001)^{0.5}$, a=2, b=2.33, $K_1$=2.06, $K_2$=2.09, $\tau$ =30. Inverse synchronization $y_1$=-$y_2$: Time series of laser system $y_1$ intensity fluctuations is shown: C is the cross-correlation coefficient between the laser systems $y_1$ and $y_2$ intensity fluctuations. Dimensionless units.

FIG.3. Numerical simulation of two delay coupled laser systems, Eqs.(9-12) for $\varepsilon$ =$(0.001)^{0.5}$, a=2, b=2.33, $K_1$=2.06, $\tau$ =30. Cross-correlation coefficient between the laser systems $y_1$ and $y_2$ intensity fluctuations vs. the coupling constant $K_2$. Dimensionless units.

Fig.1

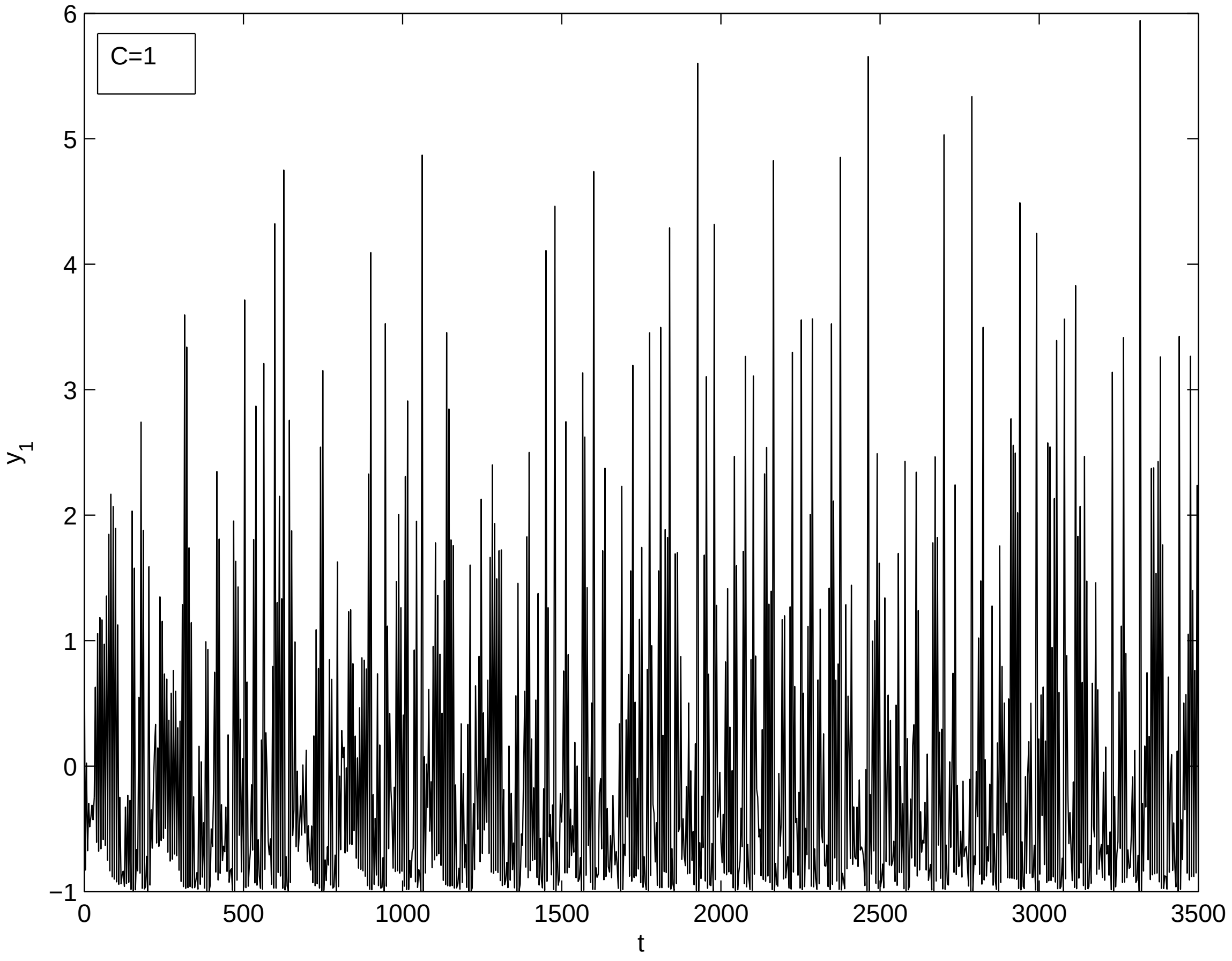

Fig.2

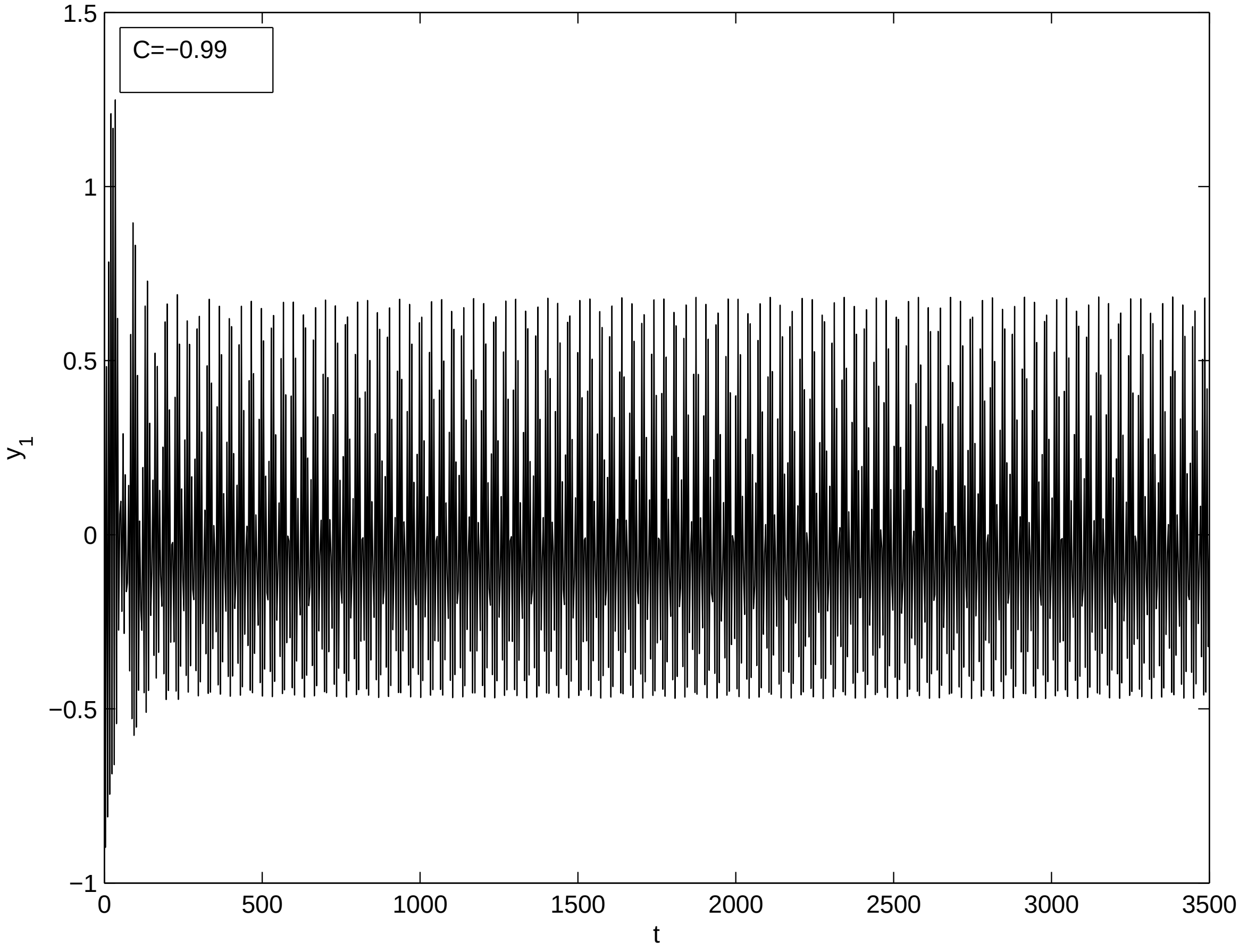

Fig.3

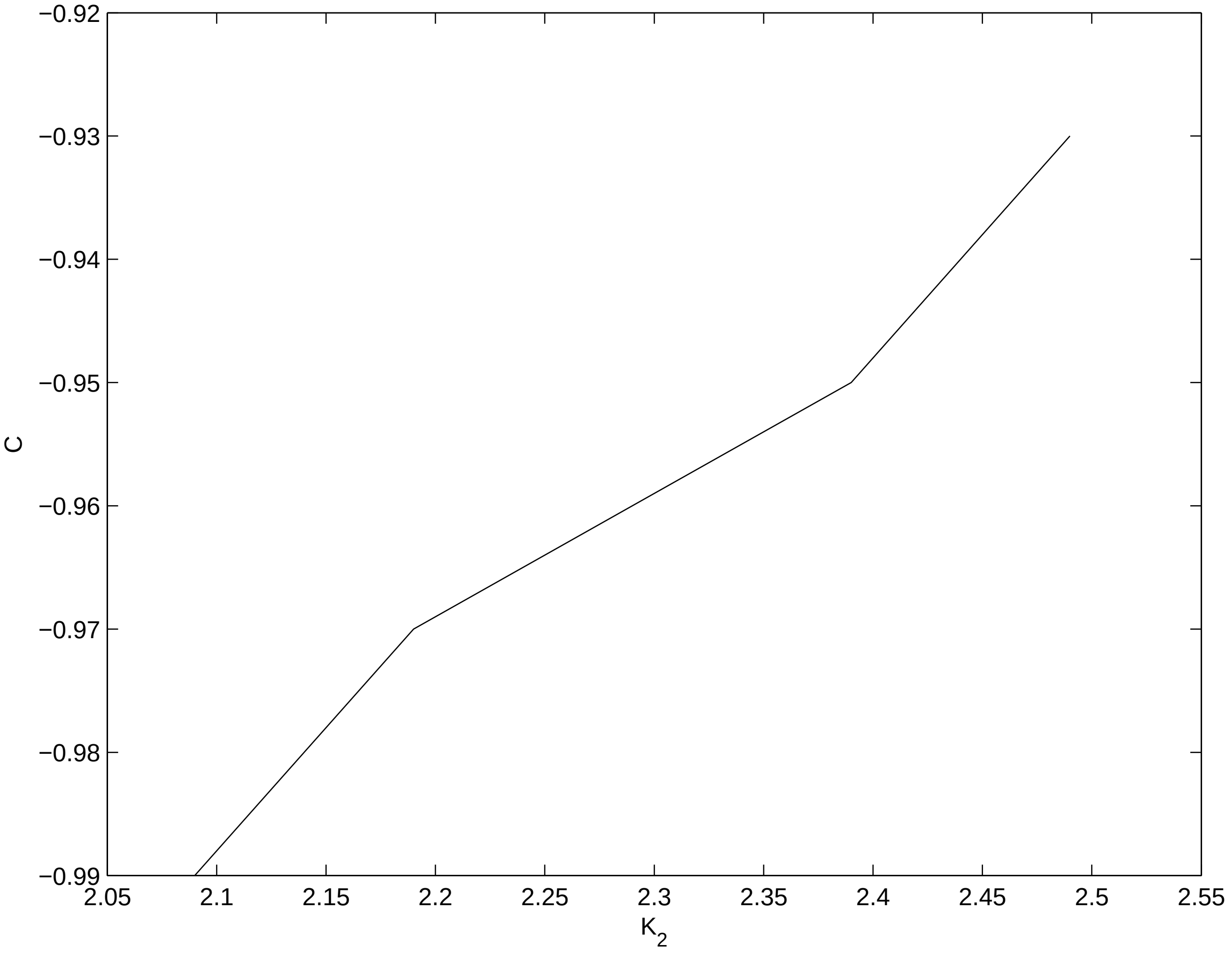